\documentclass[11pt,letterpaper]{article}
\usepackage{hyperref}
\usepackage{latexsym}
\usepackage{amssymb,amsfonts,amsmath}
\usepackage{graphicx}
\usepackage{indentfirst}
\usepackage{amsmath}
\usepackage{amsfonts}
\usepackage{amssymb}%
\usepackage{comment}
\usepackage{mathtools}
\usepackage{amsthm}
\usepackage{enumerate}
\usepackage{extarrows}
\usepackage{relsize}

\usepackage[cmtip,all]{xy}
\usepackage{verbatim}
\usepackage{tikz-cd}
\usepackage{mathtools}
\usepackage{mathrsfs}
\usepackage{geometry}

\DeclareMathAlphabet{\mathpzc}{OT1}{pzc}{m}{it}

\newcommand{\beq}{\begin{equation}}
\newcommand{\eeq}{\end{equation}}
\newcommand{\bear}{\begin{eqnarray}}
\newcommand{\eear}{\end{eqnarray}}

\ifx\pdfoutput\undefined
\else
\fi
\hypersetup{colorlinks=false,bookmarksopen,bookmarksnumbered,citecolor=blue,
pdfstartview=FitH}

\parskip=\medskipamount

\def\a{\alpha}

\def\b{\beta}

\def\g{\gamma}

\newcommand{\be}{\begin{equation}}
\newcommand{\ee}{\end{equation}}
\newcommand{\bea}{\begin{eqnarray}}
\newcommand{\eea}{\end{eqnarray}}

\newcommand{\ba}{\begin{array}}
\newcommand{\ea}{\end{array}}

\def\double #1{#1{\hbox{\kern-2pt $#1$}}}

\newcommand{\bsubeq}{\begin{subequations}}
\newcommand{\esubeq}{\end{subequations}}

\newcommand{\virgolette}{``}

\theoremstyle{definition}

\theoremstyle{plain}

\begin{document}

\begin{titlepage}
\begin{flushright}
\par\end{flushright}
\vskip 0.5cm
\begin{center}
\textbf{\LARGE \bf Power to Integral Forms}

\vskip 1cm

\large {\bf C.~A.~Cremonini}$^{~a,b,}$\footnote{carlo.alberto.cremonini@gmail.com} and 
\large {\bf P.~A.~Grassi}$^{~b, c,d,}$\footnote{pietro.grassi@uniupo.it},

\vskip .5cm {
\small
\centerline{$^{(a)}$ \it Galileo Galilei Institute for Theoretical Physics,}
\centerline{\it Largo Fermi 2, 50125, Firenze, Italy}
\centerline{$^{(b)}$ \it INFN, Sezione di Torino}
\centerline{\it via P.~Giuria 1, 10125 Torino, Italy} 
\centerline{$^{(c)}$
\it Dipartimento di Scienze e Innovazione Tecnologica (DiSIT),} 
\centerline{\it Universit\`a del Piemonte Orientale,} 
\centerline{\it viale T.~Michel, 11, 15121 Alessandria, Italy}
\centerline{$^{(d)}$
\it Arnold-Regge Center,}
\centerline{\it 
via P.~Giuria 1,  10125 Torino, Italy}
}
\end{center}

\begin{abstract}
\noindent 
A novel reformulation of D=4, N=1 supergravity action in the language of 
integral forms is given. We illustrate the construction of the Berezinian in the 
supergeometric framework, providing a useful dictionary between mathematics and 
physics. We present a unified framework for Berezin-Lebesgue integrals for 
functions and for integral forms. As an application, we discuss Volkov-Akulov 
theory and its coupling to supergravity from this new perspective. 
 \end{abstract}

\vfill{}
\vspace{1.5cm}
\end{titlepage}
\newpage\setcounter{footnote}{0}

\section{Introduction} \setcounter{equation}{0}

It has been always debated whether the superspace approach is a suitable and convenient 
framework to supergravity \cite{Wess:1977fn,VanNieuwenhuizen:1981ae}. 
Certainly, superspace is a powerful tool to deal with rigid supersymmetric theories 
(\cite{Wess:1992cp,GGRS,Buchbinder:1998qv,cube}) since it permits a compact formulation maintaining all supersymmetries manifest 
(in the case of auxiliary fields); these advantages extend beyond the classical approximation for computing 
loops and to analyse the renormalization of those models. Nonetheless, the geometric roots of superspace have not been fully exploited (see for example \cite{gates,grassi-castellani}) in the case of supergravity where the geometry of spacetime is entangled with the local supersymmetry. This is  essentially due to the lack of a complete knowledge of 
the geometry of differential forms on supermanifolds. 

Recently, based on some mathematical work 
\cite{Manin,Catenacci:2010cs,Witten:2012bg}, 
it emerged a new way to formulate supersymmetric and supergravity theories based on the concept of integral forms 
with the Picture Changing Operators (PCO) in the target space as crucial ingredients. 
The merit of this translation has to be ascribed to N. Berkovits in \cite{Berkovits:2004px}, since he noticed 
that, in the pure spinors formulation of string theory, some novel PCO's insertions 
have to be used to compute higher genus amplitudes. 
The geometrical description of those insertions goes back to seminal papers by Belopolosky \cite{Belo}  where the author introduced the use of integral forms into a more physical framework. Afterwards, one of the authors and G. Policastro started to develop a formalism to deal with those target space PCO's and 
applying it to super Chern-Simons theory \cite{Grassi:2004tv}. In \cite{grassi-castellani,Grassi:2016apf}, the authors clarify how to use the formalism to construct the action from a given Lagrangian and how to derive the correct equations of motion. 

The PCO's are related to the embedding of a bosonic submanifold into a supermanifold and this, in mathematical language, is 
known as the {\it Poincar\'e dual} to the embedding. Its  properties are $d$-closure, non-exactness and for any variation of the 
embedding in the same homology, the Poincar\'e dual varies by exact terms. 

Nonetheless, some interesting issues remain open. In particular, one can wonder 
how to translate the formalism into curved spaces and how to apply it to 
supergravity. It is clear that differential forms simplify the construction of diffeomorphism invariant theories and 
their observables \cite{cube}. The purpose is then to extend these features to 
supergravity. Here, we consider generic supervielbeins and spin connection and we show how the computations can be performed. 
Finally, we rewrite the supergravity action in a very compact and elegant way. Lastly, we apply the formalism to Volkov-Akulov theory in 
the flat and curved manifold showing its geometrical root. 

The paper is organized as follows. 
At first, we derive the volume form ${\rm Vol}^{(4|4)}$ in a $(4|4)$ dimensional supermanifold (we are interested in some applications to $N=1,D=4$ supergravity, however, all the results are easily generalizable any dimension), proving that it transforms as the Berezinian. Then, we formulate
 the integration on superspace using Dirac delta functions of differential operators. It is straightforward 
to check how these objects transform under the reparametrizations and, in addition, we put on the same level both the Berezin integration 
and the conventional Lebesgue-Riemann one. Then, along the same line, we discuss the integration measure 
on the tangent bundle (more precisely, on parity reversed tangent bundle $\Pi T ( \mathcal{SM} )$) of the supermanifold. 
The combined measure is manifestly invariant (see \cite{Witten:2012bg} and \cite{cordes}) under change of coordinates. Finally, 
we collect these ingredients to compute explicit expressions starting from integral forms or from superforms. We apply the 
formalism to $D=4, N=1$ supergravity in superspace and to Volkov-Akulov theory. In this context, we show how to couple it to 
supergravity fields and the role of PCO's depending on the Goldstino $\Lambda^\a$. In app. A and B we give some 
technical details and derivations. 

\section{SuperVielbeins and the Volume Form}

We consider $D=4$ $N=1$ superspace parametrized by the coordinates $(x^m, \theta^\mu)$ with 
$m=0,1,2,3$ and $\mu=1,2,3,4$. We use the Majorana representation for the spinors $\theta^\mu$ using 
real coordinates. We use $a=0,1,2,3$ and $\alpha =1,2,3,4$, for flat indices of the target space with  
the supercharges defined as
 $Q_\a = \partial_\a + (\gamma^a \theta)_\alpha \partial_a$ and $P_a = \partial_a$. 

The supervielbeins $E^a$ and $E^\alpha$ are $(1|0)$-superforms with values in the tangent space 
and they can be decomposed on the basis $(dx^m, d\theta^\alpha)$ as
\begin{eqnarray}
\label{exC}
E^a = E^a_m \, dx^m + E^a_\mu d\theta^\mu\,, ~~~~~~~~~~~
E^\alpha = E^\alpha_m \, dx^m + E^\alpha_\mu d\theta^\mu\,, 
\end{eqnarray}
We can write
\begin{eqnarray}
\label{exCA}
T^a =\nabla E^a - E^\alpha \gamma^a_{\alpha\beta} E^\beta\,, ~~~~~~
T^\alpha = \nabla E^\alpha\,. 
\end{eqnarray}
with $T^a, T^\a$ the vectorial and spinorial parts of the super torsion, respectively. 
The covariant derivatives are defined as $\nabla E^a = d E^a + \varpi^a_{~b} E^b$ and $\nabla E^\alpha = d E^\alpha + \frac14 \varpi_{ab} 
\gamma^{ab, \alpha}_{\beta} E^\beta$ where $\varpi^{ab}$ is the spin connection. The equations 
$\varpi^\alpha_{~\beta} = 
\frac14 (\gamma^{ab})^\alpha_{~\beta} \varpi_{ab}$ relate the spinorial representation with the vector representation. 
In the following, even if we will not explicitly use it, we set $T^a=0$ in order to fix the spin connection $\varpi_{ab}$ 
in terms of $(E^a, E^\a)$. In general, for non-vanishing dynamical supergravity fields, we cannot also set $T^\a =0$. 

The supermatrix 
\begin{eqnarray}
\label{exD}
\mathbb{E} = \left(
\begin{array}{cc}
 E^a_m  &   E^a_\mu    \\
  E^\alpha_m  &   E^\alpha_\mu 
\end{array}
\right)
\end{eqnarray}
is the well-known supervielbein appearing in supegravity. In the case of flat space, it reads 
\begin{eqnarray}
\label{exDA}
\mathbb{E}_{flat}\equiv (V^a, \psi^\alpha) = \left(
\begin{array}{cc}
 \delta^a_m  &   (\gamma^a\theta)_\mu    \\
  0  &   \delta^\alpha_\mu 
\end{array}
\right) \ ,
\end{eqnarray}
and the fluctuations around $\mathbb{E}_{flat}$ in \eqref{exDA} are identified with the dynamical vielbein and gravitino. 
Given a supermatrix, we can define the invariant superfields  
\begin{eqnarray}
\label{exE}
{\rm Ber}(\mathbb{E}) = \frac{\det(E^a_m - E^a_\mu (E^{-1})^\mu_{~\alpha} E^\alpha_m)}{\det E^\alpha_\mu} = 
 \frac{\det E^a_m }{\det (E^\alpha_\mu  - E^\alpha_m (E^{-1})^m_{~a} E^a_\mu )} \ ,
\end{eqnarray}
which are the two equivalent expressions of the \emph{superdeterminant}, well-defined when $\det E^a_m \neq 0$ and 
$\det E^\a_\mu \neq 0$. 

The volume form is the integral form with form-number $4$ and picture-number $4$ (see appendix A for an introduction to integral forms and related notations):
\begin{eqnarray}
\label{exF}
{\rm {Vol}}^{(4|4)} = \epsilon^{a_1 \dots a_4 } \delta(E^{a_1})_\wedge \dots {}_\wedge\delta(E^{a_4}) \epsilon^{\alpha_1 \dots \alpha_4} \delta(E^{\alpha_1})_\wedge \dots {}_\wedge\delta(E^{\alpha_4})   \ ,
\end{eqnarray}
which corresponds to a top form in supergeometry. 
It is closed, $d{\rm Vol}^{(4|4)}=0$, and non-exactness depends on the $(4|4)$ supermanifold on which it is defined (for example, if the supermanifold is compact, the top form is non-exact). The cohomological properties can be easily checked by applying the differential $d$, the Leibniz rule for $\nabla$ acting on the supervielbeins $E^\a, E^\alpha$ and the 
distributional equations $E^\a \delta(E^\a)=0$.

In addition, the volume form ${\rm Vol}^{(4|4)}$ is invariant with respect to Lorentz transformations  
\begin{eqnarray}
\label{exFA}
\delta E^a = \Lambda^a_{~b} E^b\,, ~~~~~~~~~
\delta E^\alpha = \frac14 \Lambda_{ab} (\gamma^{ab})^\alpha_{~\beta} E^\beta \ ,
\end{eqnarray}
then, more precisely, it belongs to an equivariant cohomology. 

Since the $E^a$'s are anticommuting (bosonic 1-forms), we can replace the first delta's with their arguments
\begin{eqnarray}
\label{exG}
{\rm Vol}^{(4|4)} = \epsilon_{a_1 \dots a_4}E^{a_1}_\wedge \dots {}_\wedge E^{a_4} \epsilon^{\alpha_1 \dots \alpha_4} \delta(E^{\alpha_1})_\wedge \dots {}_\wedge \delta(E^{\alpha_4}) \ .
\end{eqnarray}
The same cannot be done for $E^\alpha$'s, since they are commuting. Notice that the $\delta(E^\a)$ does not transform as a tensor with respect to 
change of parametrization and therefore the index $\alpha$ is not a conventional covariant index summed with the 
Levi-Civita tensor $ \epsilon^{\alpha_1 \dots \alpha_4}$; the latter serves only to keep track of the order of the deltas, because of their anticommutation relations 
$\delta(E^\a)_\wedge \delta(E^\beta) = - \delta(E^\beta)_\wedge \delta(E^\a)$. Nonetheless, expression \eqref{exG} is invariant under reparametrizations. 
One has to pay some attention to deal with those ``covariant" expressions (in \cite{Cremonini:2019aao}, we give a 
set of rules that indicate how to deal with single delta's). 
 
Inserting expressions (\ref{exC}) into (\ref{exG}), 
using the properties of the oriented delta's and of the 1-forms, we get 
\begin{eqnarray}
\label{exH}
{\rm Vol}^{(4|4)}  = {\rm Ber}(\mathbb{E}) \, 
\epsilon_{m_1 \dots m_4 }dx ^{m_1}_\wedge \dots {}_\wedge dx^{m_4} \epsilon^{\mu_1 \dots \mu_4} \delta(d\theta^{\mu_1})_\wedge \dots {}_\wedge \delta(d\theta^{\mu_4}) \ ,
\end{eqnarray}
where the overall factor is the superdeterminant of $\mathbb{E}$ \eqref{exF}, as expected. 

\section{Vector Fields and the Berezinian Integration}

Let us move to vector fields. The covariant 
vector fields are denoted by $(\nabla_a, \nabla_\alpha)$ and they can be decomposed on the basis $(\partial_m, \partial_\mu)$ as
\begin{eqnarray}
\label{exL}
\nabla_a = E_a^m \partial_m + E_a^\mu \partial_\mu\,, ~~~~~~
\nabla_\alpha =  E_\alpha^m \partial_m + E_\alpha^\mu \partial_\mu\,,
\end{eqnarray}
satisfying the usual commutation relations 
\begin{eqnarray}
\label{exLA}
&&[\nabla_a, \nabla_b] = T_{ab}^c \nabla_c + T_{ab}^\gamma \nabla_\gamma + R_{ab}\,,  \nonumber \\
&&[\nabla_a, \nabla_\beta] = T_{a\beta}^c \nabla_c  + T_{a \beta}^\gamma \nabla_\gamma + R_{a\beta}\,, \nonumber \\ 
&&\{\nabla_\alpha, \nabla_\beta\} = T_{\alpha\beta}^c \nabla_c  + T_{\alpha \beta}^\gamma \nabla_\gamma + R_{\alpha\beta}\,, 
\end{eqnarray}
where $T$ are the components of the torsion and $R$ are the components of the curvature. Choices on the components 
of $T$ and $R$ are usually made to reduce redundant and unphysical degrees of freedom. 
The relations between superderivatives and supervielbeins are
\begin{eqnarray}
\label{exM}
\iota_{\nabla_a} E^b = \delta_a^b\,, ~~~~~~
\iota_{\nabla_a} E^\beta = 0 \,, ~~~~~~
\iota_{\nabla_\alpha} E^b = 0\,, ~~~~~~
\iota_{\nabla_\alpha} E^\beta = \delta_\alpha^\beta\,, ~~~~~~
\end{eqnarray}
which can be expressed in components as 
\begin{eqnarray}
\label{exN}
&&E_a^m E_m^b + E_a^\mu E_\mu^b =  \delta_a^b\,, ~~~~~~~~
E_a^m E_m^\beta + E_a^\mu E_\mu^\beta =  0\,, ~~~~~~~~\nonumber \\
&&E_\alpha^m E_m^b + E_\alpha^\mu E_\mu^b =  0\,, ~~~~~~~~
E_\alpha^m E_m^\beta + E_\alpha^\mu E_\mu^\beta = \delta_\alpha^\beta\,.
\end{eqnarray}
Therefore, they are not independent and, solving these equations, they can be easily related. If 
we denote the supermatrix built from the components of $(\nabla_a, \nabla_\alpha)$ as 
\begin{eqnarray}
\label{exDAinverse}
\mathbb{E}^{-1} = \left(
\begin{array}{cc}
 E_a^m  &   E_a^\mu    \\
  E_\alpha^m  &   E_\alpha^\mu 
\end{array}
\right) \ ,
\end{eqnarray}
we have $\mathbb{E} \mathbb{E}^{-1} = 1$, which is the identity matrix on $(n|m)$ vector space. Moreover, it follows that
\begin{eqnarray}
\label{exDB}
{\rm Ber }(\mathbb{E}^{-1}) = \frac{1}{{\rm Ber}(\mathbb{E})} \ .
\end{eqnarray}

Let us introduce now the operator 
\begin{eqnarray}
\label{veA}
\mathlarger{\mathlarger \Xi} &=&  \epsilon_{a_1 \dots a_4 } \delta(i \nabla_{a_1})\dots \delta(i \nabla_{a_4}) 
\epsilon_{\alpha_1 \dots \alpha_4} \delta(\nabla_{\alpha_1}) \dots  \delta(\nabla_{\alpha_4})  \nonumber \\
 &=&  \epsilon_{a_1 \dots a_4 } \delta(i \nabla_{a_1}) \dots \delta(i \nabla_{a_4}) \epsilon^{\alpha_1 \dots \alpha_4} \nabla_{\alpha_1} \dots \nabla_{\alpha_4} \ ,
\end{eqnarray}
where we added the $i$ factors in order to deal with hermitian operators. Since $\nabla_{\a}$ are anticommuting vector fields we can 
replace $\delta(\nabla_{\alpha})$ with its argument $\nabla_{\alpha}$, while for the commuting vector fields $\nabla_a$ this cannot be done, in analogy to \eqref{exG}. 

As above, the expression $\delta(i \nabla_{a})$ is not a tensor and some care has to be paid. Now, we study how this operator works on test functions $f(x, \theta)$: first we consider the flat vector field $\partial_m$, then we translate the result to the curved case $ \delta(i \nabla_{a})$. 

Using the integral representation for the Dirac delta's, we have 
\begin{eqnarray}
\label{veBintegral}
\delta(i \partial_m) f(x_1, \dots, x_4) &=& \int_{-\infty}^\infty dt e^{i t (i \partial_m)} f(x_1, \dots, x_4) \nonumber \\
&=& 
\int_{-\infty}^\infty dt f(x_1, ..., x_{m-1}, x_m - t,x_{m+1}, ...x_4) 
\nonumber \\
&=& 
 \int_{-\infty}^\infty dt' f(x_1, ..., x_{m-1}, t',x_{m+1},... x_4)  
\end{eqnarray}
acting on a test function $f$. Notice that this formulation is consistent with Stokes' theorem (we will comment further on this in the following sections):
\begin{eqnarray}
\label{veB}
\delta(i \partial_m) \partial_m f(x_1, \dots, x_4) =0 \ ,
\end{eqnarray}
which on one hand follows from the (formal) distributional property $\delta \left( \partial_m \right) \partial_m = 0$ and, on the other hand, 
from fundamental theorem of calculus, for any test function $f$. 
In addition, using integration by parts, it is 
easy to show that 
\begin{eqnarray}
\label{veC}
\delta'(i \partial_m) \partial_m f(x_1, \dots, x_4) = - i \delta(i \partial_m) f(x_1, \dots, x_4) \ ,
\end{eqnarray}
which are the usual properties of Dirac delta's. It also follows the well-known property 
$\delta(i \partial_m) \delta(x^m) = 1$ (not summed on indices $a$). 
Putting all together, we get the final expression
\begin{eqnarray}
\label{veD}
\epsilon^{m_1 \dots m_4} \delta(i\partial_{m_1}) \dots \delta(i\partial_{m_4})
f(x_1, \dots, x_4) = 
\int_{-\infty}^\infty dt_1 \dots dt_n  f(t_1, \dots, t_4) \ ,
\end{eqnarray}
for the bosonic part of the integration, 
and 
\begin{eqnarray}
\label{veE}
&&\epsilon^{m_1 \dots m_n} \delta(\partial_{m_1}) \dots \delta(\partial_{m_4})
\epsilon^{\alpha_1 \dots \mu_4} \delta(\partial_{\mu_1}) \dots \delta(\partial_{\mu_4}) 
f(x_1, \dots, x_n, \theta_{\mu_1}, \dots, \theta_{\mu_4}) =\nonumber \\
&&= 
\int \epsilon^{m_1 \dots m_4} dt_{m_1} \dots dt_{m_4} \epsilon^{\mu_1 \dots \mu_4} \partial_{\mu_1} 
\dots \partial_{\mu_4} f(t_1, \dots, t_4, \theta_{\mu_1}, \dots, \theta_{\mu_4}) \ ,
\end{eqnarray}
for the superspace version. 
The expression $d\mu =  \epsilon^{m_1 \dots m_4} dt_{m_1} \dots dt_{m_4} \epsilon^{\mu_1 \dots \mu_4} 
\partial_{\mu_1} \dots \partial_{\mu_4} $ is the Berezin integral measure (see \cite{Cacciatori:2020hcm} and \cite{noja2020note} for a discussion from a more rigorous point of view). 
The crucial point here is to observe that placing on the same level both the derivatives $\partial_m$ and the 
super derivatives $\partial_\mu$ we deduce that indeed the Berezin integral on a Grassmann coordinate is 
a derivative. This is due to the fact that a Dirac delta of a Grassmann argument is proportional to the argument itself. 

Let us study the transformation properties of (\ref{veA}). For that, we insert the covariant derivatives 
$\nabla_a, \nabla_\alpha$ and we use (\ref{exL}). 
In the derivation, we use the properties of Dirac delta's and the antisymmetrization with respect to $a$ and $\alpha$ indices (we write the derivation for a generic dimension $(n|m)$):

\begin{eqnarray}\label{veF}
\Xi &&=
\epsilon^{a_1 \dots a_n} \delta(i\nabla_{a_1}) \dots \delta(i\nabla_{a_n})
\epsilon^{\alpha_1 \dots \alpha_n} \delta(i\nabla_{\alpha_1}) \dots \delta(i\nabla_{\alpha_m}) \nonumber \\
&&
=\epsilon^{a_1 \dots a_n} \delta(i\nabla_{a_1}) \dots \delta(i\nabla_{a_n})
\epsilon^{\alpha_1 \dots \alpha_n}\nabla_{\alpha_1} \dots \nabla_{\alpha_m}\nonumber \\
&&
=\epsilon^{a_1 \dots a_n} \delta\left(E_{a_1}^{m_1} \partial_{m_1} + E_{a_1}^{\mu_1} \partial_{\mu_1}\right) \dots 
\delta\left(E_{a_n}^{m_n} \partial_{m_n} + E_{a_n}^{\mu_n} \partial_{\mu_n}\right)\times \nonumber \\
&&
\epsilon^{\alpha_1 \dots \alpha_m} 
\left(E_{\alpha_1}^{m_1} \partial_{m_1} + E_{\alpha_1}^{\mu_1} \partial_{\mu_1}\right) \dots 
\left(E_{\alpha_m}^{r_m} \partial_{r_m} + E_{\alpha_m}^{\mu_m} \partial_{\mu_m}\right) \nonumber \\
&&
=\epsilon^{a_1 \dots a_n} (E^{-1})^{a_1}_{m_1} \dots (E^{-1})^{a_n}_{m_n} 
\delta\left( \partial_{m_1} + (E^{-1})^{a_1}_{m_1} E_{a_1}^{\mu_1} \partial_{\mu_1}\right) \dots 
\delta\left( \partial_{m_n} + (E^{-1})^{a_n}_{m_n} E_{a_n}^{\mu_n} \partial_{\mu_n}\right) \times \nonumber \\
&&
\epsilon^{\alpha_1 \dots \alpha_m} 
\left(E_{\alpha_1}^{m_1} \partial_{m_1} + E_{\alpha_1}^{\mu_1} \partial_{\mu_1}\right) \dots 
\left(E_{\alpha_m}^{r_m} \partial_{r_m} + E_{\alpha_m}^{\mu_m} \partial_{\mu_m}\right) = \nonumber \\
&&
=\epsilon^{a_1 \dots a_n}  (E^{-1})^{a_1}_{m_1} \dots (E^{-1})^{a_n}_{m_n}  
\delta\left( \partial_{m_1} + (E^{-1})^{a_1}_{m_1} E_{a_1}^{\mu_1} \partial_{\mu_1}\right) \dots 
\delta\left( \partial_{m_n} + (E^{-1})^{a_n}_{m_n} E_{a_n}^{\mu_n} \partial_{\mu_n}\right) \times \nonumber \\
&&
\epsilon^{\alpha_1 \dots \alpha_m} 
\left(- E_{\alpha_1}^{m_1}(E^{-1})^{a_1}_{m_1} E_{a_1}^{\mu_1} \partial_{\mu_1} + E_{\alpha_1}^{\mu_1} \partial_{\mu_1}\right) \dots 
\left(- E_{\alpha_m}^{r_m}(E^{-1})^{a_m}_{r_m} E_{a_m}^{\mu_m} \partial_{\mu_m} + E_{\alpha_m}^{\mu_m} \partial_{\mu_m}\right) \nonumber \\
&&
=\epsilon^{a_1 \dots a_n} (E^{-1})^{a_1}_{m_1} \dots (E^{-1})^{a_n}_{m_n} 
\delta\left( \partial_{m_1}\right) \dots 
\delta\left( \partial_{m_n}\right) \times \nonumber \\
&&
\epsilon^{\alpha_1 \dots \alpha_m} 
\left(- E_{\alpha_1}^{m_1}(E^{-1})^{a_1}_{m_1} E_{a_1}^{\mu_1} + E_{\alpha_1}^{\mu_1} \right) \partial_{\mu_1}\dots 
\left(- E_{\alpha_m}^{r_m}(E^{-1})^{a_m}_{r_m} E_{a_m}^{\mu_m} + E_{\alpha_m}^{\mu_m} \right) \partial_{\mu_m} \nonumber \\
&&
=\frac{\det\left(E_{\alpha}^{\mu} - E_{\alpha}^{m}(E^{-1})^{b}_{m} E_{b}^{\mu}\right)}{\det 
E_{a}^{m}} \epsilon^{m_1 \dots m_n} \delta\left( \partial_{m_1}\right) \dots 
\delta\left( \partial_{m_n}\right) \epsilon^{\mu_1 \dots \mu_m} \partial_{\mu_1}\dots \partial_{\mu_m}  \nonumber \\
&&
={\rm Ber}(\mathbb{E}^{-1}) \, \epsilon^{m_1 \dots m_n} \delta\left(i \partial_{m_1}\right) \dots 
\delta\left( i\partial_{m_n}\right) \epsilon^{\mu_1 \dots \mu_m} \partial_{\mu_1}\dots \partial_{\mu_m} \ .
\end{eqnarray}
Then, we obtained
\begin{eqnarray}
\label{veG}
\Xi = {\rm Ber}(\mathbb{E}^{-1}) \, \epsilon^{m_1 \dots m_n} \delta\left(i \partial_{m_1}\right) \dots 
\delta\left(i \partial_{m_n}\right) \epsilon^{\mu_1 \dots \mu_m} \partial_{\mu_1}\dots \partial_{\mu_m} \ ,
\end{eqnarray}
verifying that the formal integration symbol transforms as the inverse of the superdeterminant of the Jacobian of a generic change of coordinates.
Therefore, it transforms in the opposite way as ${\rm Vol}^{(n|m)}$ in (\ref{exH}). The non-linear structure 
emerges from the redefinition of the derivatives in terms of covariant derivatives. The transformation properties of $\Xi$ 
embodies the original proof of Berezin \cite{Berezin} for integrals in superspace. In  
\cite{noja2020note}, the transformation properties of $d\mu$ are proved in the cohomology of Koszul complex, 
here the computation is explicit. Then, given a scalar superfield $\Phi(x, \theta)$ we set  
\begin{eqnarray}
\label{veGA}
\Xi[\Phi(x, \theta) ] = \int  {\rm Ber}(\mathbb{E}) \Phi(x, \theta) \ ,
\end{eqnarray}
which is the usual Berezin-Lebesgue integral used in superspace formulations. 
Again we point out that $\delta(d\theta^\mu)$ and $\delta(i \partial_m)$ are not tensors and 
only the products of all those operators transform as pseudodensities. In absence of further 
geometric structures (such as a metric or a globally-defined vector field) it is hard to give 
a precise mathematical meaning to a single $\delta(d\theta)$ from tensorial point of view. 
A detailed analysis of pseudoforms, namely those forms with non-maximal and non-zero picture number, 
in a physical interesting model as been pursued in \cite{Cremonini:2019aao}. 

We recall the duality relations between forms and vector fields
\begin{eqnarray}
\label{foA}
\iota_{\partial_a} dx^b = \langle dx^b, \partial_a \rangle =  \delta^b_a\,, ~~~~~~~ , ~~~~~~\iota_{\partial_\alpha} d\theta^\beta = \langle d\theta^\beta, \partial_\alpha\rangle  =\delta^\beta_\alpha\,. 
\end{eqnarray}
If one considers $\delta(d\theta)$, the analogous dual operation is 
\begin{eqnarray}
\label{foB}
\delta(\iota_{\partial_\alpha}) \delta(d\theta^\alpha) = 1 ~~~ ({\rm not~summed~over~}\alpha) \ .
\end{eqnarray}
Of course, $\delta(dx^a) = dx^a$ and, in the same way $\delta(\iota_{\partial_a})  = \iota_{\partial_a}$. 
We report also the following results 
\begin{eqnarray}
\label{foC}
\delta(\partial_\alpha) \delta(\theta^\beta) = \partial_\alpha \theta^\beta = \delta_\alpha^\beta\,, ~~~~~
\delta(\partial_a) \delta(x^a) = 1  ~~~ ({\rm not~summed~over~}a)
\,. 
\end{eqnarray}
as can be easily checked by the definitions given above. 

\section{Integration on the Tangent Bundle}

In the previous section we derived the expression $\Xi$ which describes the integration on the supermanifold (in \cite{Witten:2012bg} this is denoted 
by the symbol $[dx d\theta]$). Here, we would like to complete it to the entire cotangent supermanifold $\Pi T \left( {\mathcal SM} \right)$, (in Witten's notation $[dx d\theta d(dx) d(d\theta)]$). The parity-changing functor $\Pi$ takes into account that 
we are not integrating on the real tangent bundle  $T \left( {\mathcal SM} \right)$, but rather on the supermanifold where $(x^a, d\theta^\a)$ are commuting coordinates, while $(dx^a, \theta^\a)$ are anticommuting ones. 
 
This can be done by translating the differential operators $\nabla_a, \nabla_\a$ acting on $(x^a, \theta^\alpha)$ 
into the contractions $\iota_{\nabla_a}, \iota_{\nabla_\alpha}$ acting on differential 1-forms $(dx^a, d\theta^\alpha)$. We can then define 
\begin{eqnarray}
\label{totA}
\mathlarger{\mathlarger \Xi^*} &=& \, \epsilon^{a_1 \dots a_n} \delta(\iota_{\nabla_{a_1}}) \dots  \delta(\iota_{\nabla_{a_n}}) 
\epsilon^{\a_1 \dots \a_n} \delta(i\iota_{\nabla_{\a_1}}) \dots  \delta(i\iota_{\nabla_{\a_m}})  
\nonumber \\
&=& 
 \, \epsilon^{a_1 \dots a_n} \iota_{\nabla_{a_1}} \dots  \iota_{\nabla_{a_n}} 
\epsilon^{\a_1 \dots \a_n} \delta(i \iota_{\nabla_{\a_1}}) \dots  \delta(i \iota_{\nabla_{\a_m}}) \ ,
\end{eqnarray}
where since the contractions $ \iota_{\nabla_{a}}$ are odd differential operators, 
we removed the Dirac delta symbol. On the other side, we can not remove the Dirac delta symbol 
for the contraction along the anticommuting vector fields $\nabla_\a$, nonetheless, we 
can use their integral representation
\begin{eqnarray}
\label{totB}
 \delta(i\iota_{\nabla_{\a}}) = \int_{-\infty}^\infty dt e^{- t \iota_{\nabla_\alpha}} \ ,
\end{eqnarray}
so that they can act on $\delta(d\theta^\a)$. 
It is easy to prove that $\Xi^*$  transforms as ${\rm Ber}({\mathbb E})$ with the same computation as in (\ref{veF}). 
The expression for the measure on full $\Pi T \left( {\mathcal SM} \right)$ is finally given by 
\begin{eqnarray}
\label{totBA}
\mathcal{\mathlarger{\mathlarger \mu}} =\mathlarger{\mathlarger \Xi} \mathlarger{\mathlarger \Xi^*}= \prod_{i=1}^n
\delta(i\nabla_{a_i}) \iota_{\nabla_{a_i}} \prod_{j=1}^m 
\nabla_{\alpha_j} \delta(i \iota_{\nabla_{\a_j}}) \ ,
\end{eqnarray}
and it corresponds to $\int[d^nxd^m\theta d^n(dx) d^m(d\theta)]$ using the notation of \cite{Witten:2012bg}. The new measure 
is invariant under any change of coordinates since the transformation of $\Xi$ compensates the transformation 
of $\Xi^*$. 

Given now a top integral form $\omega^{(n|m)}$ which can be expressed as 
\begin{eqnarray}
\label{totBDA}
\hspace{-.5cm} \omega^{(n|m)} = \Phi(x,\theta)   {\rm Vol}^{(n|m)} =  \Phi(x,\theta)   \epsilon^{a_1 \dots a_n } \delta(E^{a_1})_\wedge \dots {}_\wedge\delta(E^{a_n}) \epsilon^{\alpha_1 \dots \alpha_m} \delta(E^{\alpha_1})_\wedge \dots {}_\wedge\delta(E^{\alpha_m}) \ ,
\end{eqnarray}
where $\Phi(x,\theta) $ is an invariant superfield, we can integrate it as follows 
\begin{eqnarray}
\label{totBDB}
\mathlarger{\mathlarger \mu} \left[ \omega^{(n|m)} \right]= \mathlarger{\mathlarger \Xi} \mathlarger{\mathlarger \Xi^*} \left[
\Phi(x,\theta)    {\rm Vol}^{(n|m)} \right] = 
 \mathlarger{\mathlarger \Xi} \left[ \Phi(x,\theta)   \right] = \int  {\rm Ber}(\mathbb{E})  \Phi(x, \theta) \ .
\end{eqnarray}
Hence $\mathlarger{\mathlarger \mu}$ defines a functional on the Berezinian bundle of the supermanifold 
${\rm Ber}[\Pi T {\mathcal SM}]$. Notice that 
since $\Phi$ is a scalar, the last integral is invariant under super-reparametrizations. 

It is worth noticing that the factors appearing in the second product of \eqref{totBA} can be written 
as (for any $\alpha_j$)
\begin{eqnarray}
\label{totBB}
\nabla_{\alpha_j} \delta(i \iota_{\nabla_{\a_j}}) \rightarrow -i \left[ d, \Theta(i \iota_{\nabla_{\alpha_j}}) \right] \equiv Z_{\alpha_j} \ .
\end{eqnarray}
The proof is not straightforward, one has to replace the covariant derivatives $\nabla_{\alpha_j}$ 
with the Lie derivative ${\mathcal L}_{\nabla_{\alpha_j}}$ and this can be done because of the factor 
$\prod_{j=1}^m \delta(i \iota_{\nabla_{\a_j}})$ and the distributional equation $\iota_\a \delta(i \iota_\a) =0$.  
Then, using the Cartan formula ${\mathcal L}_{\nabla_{\alpha_j}} = d \iota_{\nabla_{\a_j}} + \iota_{\nabla_{\a_j}} d$, 
one gets the result. 
$ Z_{\alpha_j}$ is the PCO (picture lowering operator) appearing in string theory. It decreases the picture by 
removing Dirac delta's $\delta(d\theta^\a)$. It is closed, but it is not exact, since it is expressed in terms 
of a non-compact distribution $\Theta(i \iota_{\nabla_{\alpha_j}})$. Again, the easiest way to deal with the operator 
$\Theta(i \iota_{\nabla_{\alpha_j}})$ is by using its integral representation 
\begin{eqnarray}
\label{totBC}
 \Theta(i\iota_{\nabla_{\a}}) = - \int_{-\infty}^\infty dt \frac{e^{- t \iota_{\nabla_\alpha}}}{t + i \epsilon} \ ,
\end{eqnarray}
acting on the space of integral forms. The expression \eqref{totBA} can be analogously written as 
\begin{eqnarray}
\label{totBD}
\mathcal{\mathlarger{\mathlarger \mu}} =\mathlarger{\mathlarger \Xi} \mathlarger{\mathlarger \Xi^*}= \prod_{i=1}^n
\delta(i\nabla_{a_i}) \iota_{\nabla_{a_i}} \prod_{j=1}^m Z_{\alpha_j} \ .
\end{eqnarray}
The factor with the PCO $Z$'s parallels the measure both in NRS and pure spinor string theory. Higher genus computations 
require a number of PCO's equivalent to the number of moduli and of zero modes on the Riemann surface.  Notice that 
the PCO's $Z$ emerge naturally in the present geometric framework (see also \cite{Cremonini:2019aao}).

It is interesting to observe that the PCO
\begin{eqnarray}
\label{totC}
\mathbb{Y}^{(0|4)} = \epsilon_{\mu_1 \dots \mu_4} \theta^{\mu_1} \dots \theta^{\mu_4} 
\epsilon_{\mu_1 \dots \mu_4} \delta(d\theta^{\mu_1}) \dots \delta(d\theta^{\mu_4}) \ , 
\end{eqnarray}
which is closed and not exact, corresponds to the trivial  embedding of the bosonic submanifold 
${\mathcal M}^{(4)}$ into the supermanifold ${\mathcal{SM}}^{(4|4)}$, i.e., the one that sets $\theta^\a =0$ and $d\theta^\a=0$ 
for 
every $\alpha$. As discussed in App. B, this PCO represents a cohomology class and  hence its representative 
can be changed inside the same class by adding exact pieces. In App. B it is 
also discussed the curved version of $\mathbb{Y}^{(0|4)}$:
\begin{eqnarray}
\label{totCAA}
\mathbb{Y}^{(0|4)}= \epsilon_{\alpha_1 \dots \alpha_4}\iota_{\mathcal E} E^{\alpha_1} 
\dots \iota_{\mathcal E} E^{\alpha_4} \delta(E^{\alpha_1}) \dots \delta(E^{\alpha_4}) \ , 
\end{eqnarray}
where ${\mathcal E}$ is the Euler vector along the fermionic coordinates.

The PCO's $\mathbb{Y}^{(0|4)}$, contrary to $Z$, raise the picture by 4,
converting a superform $\omega^{(p|0)}$ into an integral form $\omega^{(p|4)} = \omega^{(p|0)} \wedge \mathbb{Y}^{(0|4)}$ 
which can be integrated on  $(p|4)$-dimensional submanifolds of the supermanifold. For example, if $p=4$, we get 
\begin{eqnarray}
\label{totCA}
\mathcal{\mathlarger{\mathlarger \mu}} \left[  \omega^{(4|0)} \wedge \mathbb{Y}^{(0|4)}\right] = 
\int_{{\mathcal M}^{(4)}} \left. \omega^{(4|0)} \right|_{\theta=d\theta=0} \ ,
\end{eqnarray}
reducing the expression to a conventional integral of the bosonic top form $\left. \omega^{(4|0)} \right|_{\theta=d\theta=0}$. This 
equation was discussed in pioneering works \cite{gates} and in \cite{grassi-castellani}. 
Using the PCO $\mathbb{Y}$, we can establish a corresponding operator $\mathcal{\mathlarger{\mathlarger \mu}}_B$ 
for the spacetime forms
\begin{eqnarray}
\label{totD}
\mathcal{\mathlarger{\mathlarger \mu}}_B\left[ \omega^{(4)*} \right] = \mathcal{\mathlarger{\mathlarger \mu}}\left[ \omega^{(4|0)}\wedge 
\mathbb{Y}^{(0|4)}\right] \,,
\end{eqnarray}
which then encodes the bosonic part of the integral only and where $ \omega^{(4)*} $ is the 4-form pulled-back on the bosonic submanifold. 
We finally notice that if $\omega$ is 
\begin{eqnarray}
\label{totF}
{\mathbb S}^{(4|0)} = \delta(x^{m_1}) \dots \delta(x^{m_4}) dx^{m_1}_\wedge \dots {}_\wedge dx^{m_4} \ ,
\end{eqnarray}
we have that the application of the functional $\mu$ to $\sigma^{(4|4)} = {\mathbb S}^{(4|0)} \wedge \mathbb{Y}^{(0|4)}$ gives
\begin{eqnarray}
\label{totG}
\mathcal{\mathlarger{\mathlarger \mu}}\left[ {\mathbb S}^{(4|0)} \wedge \mathbb{Y}^{(0|4)}\right] = 1 \ ,
\end{eqnarray}
where $\sigma^{(4|4)}$ is a representative of the singular cohomology discussed in Bott-Tu \cite{bott-tu}. It is worth noticing that 
\begin{eqnarray}
\label{totGA}
d( {\mathbb S}^{(4|0)} \wedge \mathbb{Y}^{(0|4)})  = d {\mathbb S}^{(4|0)} \wedge \mathbb{Y}^{(0|4)} + 
{\mathbb S}^{(4|0)} \wedge d \mathbb{Y}^{(0|4)}= 0 \ ,
\end{eqnarray}
hence $\sigma^{(4|4)}$ is closed and not exact. In particular, notice that one could write $\delta(x) dx = [d, \Theta(x)]$ which is exact, but not in the 
same distributional space. In other words, the fact that $\sigma = \delta(x) dx$ is a cohomology class is obvious by the fact that 
$\int \sigma =1$. 
The form $\mathbb{S}$ can be also viewed as the Poincar\'e dual of an 
immersion of a point into a manifold. In the same way, $\sigma^{(4|4)}$ is the Poincar\'e dual of the 
immersion of a point into a supermanifold ${\mathcal{SM}}^{(4|4)}$. The case of lines or of 
2d surfaces immersed into a supermanifold ${\mathcal{SM}}^{(m|n)}$ are discussed in \cite{Cremonini:2020mrk}. 


As is well-known, Stokes' theorem is a crucial ingredient to write consistent expressions, to derive equations of motion and 
to manipulate equations. In the present formalism we show how is proven. In this section we will avoid boundary terms, in the following we include them. 
Let us consider an integral form $\omega^{(3|4)}$ which is decomposed locally as 
\begin{eqnarray}
\label{sto0A}
\omega^{(3|4)} &=& \omega^a(x,\theta) \epsilon_{a b_1 b_2 b_3} E^{b_1} \dots E^{b_3} 
\epsilon^{\a_1 \dots \a_4} \delta(E^{\a_1}) \dots  \delta(E^{\a_4}) \nonumber \\&+& 
\omega^\a(x,\theta) \epsilon_{b_1 b_2 b_3 b_4} E^{b_1} \dots E^{b_4} 
\epsilon^{\a_1 \dots \a_4}\iota_{\nabla_\alpha}  (\delta(E^{\a_1}) \dots  \delta(E^{\a_4})) \ ,
\end{eqnarray}
where $\iota_{\nabla_\alpha} (\delta(E^{\a_1}) \dots  \delta(E^{\a_4}))$ is the derivative of the 
delta functions. 
The integral form depends on the two superfields 
$\omega^a$ and $\omega^\a$ and the indices 
on $\omega^a, \omega^\a$ are contravariant. We can recast the expression for \eqref{sto0A} in a more useful way as
\begin{eqnarray}
	\omega^{(3|4)} &=& \left[ \omega^a \left( x, \theta \right) \iota_{\nabla_a} + \omega^{\alpha} \left( x , \theta \right) \iota_{\nabla_\alpha} \right] \text{Vol}^{(4|4)} \ ,
\end{eqnarray}
where we have used the expression defined in \eqref{exG}. We can compute the differential $d=E^a \nabla_a + E^\alpha \nabla_\alpha$, use \eqref{exM} and the distributional identities to get
\begin{eqnarray}
\label{sto0B}
d \omega^{(3|4)}  =\left( \nabla_a \omega^a + \nabla_\a  \omega^\a \right) 
\epsilon_{b_1 b_2 b_3 b_4} E^{b_1} \dots E^{b_4} \epsilon^{\a_1 \dots \a_4} \delta(E^{\a_1}) \dots  \delta(E^{\a_4}) \ ,
\end{eqnarray}
which is a $(4|4)$ form. We can integrate \eqref{sto0B} and verify that its integral vanishes, modulo boundary terms:
\begin{eqnarray}
\label{stoA}
&&\mathcal{\mathlarger{\mathlarger \mu}}\left[ d \omega^{(3|4)} \right] =  
\prod_{i=1}^4
\delta(i\nabla_{a_i}) \iota_{\nabla_{a_i}} \prod_{j=1}^4 
\nabla_{\alpha_j} \delta(i \iota_{\nabla_{\a_j}}) \left[ d \omega^{(3|4)} \right] \nonumber \\
&& = 
\prod_{i=1}^4
\delta(i\nabla_{a_i}) \iota_{\nabla_{a_i}} \prod_{j=1}^4 
\nabla_{\alpha_j} \delta(i \iota_{\nabla_{\a_j}}) \left[ \nabla_a \omega^a + \nabla_\a \omega^\a\right]
\epsilon_{b_1 b_2 b_3 b_4} E^{b_1} \dots E^{b_4} \epsilon^{\a_1 \dots \a_4} \delta(E^{\a_1}) \dots  \delta(E^{\a_4})\nonumber 
\\
&& =\prod_{i=1}^4
\delta(i\nabla_{a_i}) \prod_{j=1}^4 
\nabla_{\alpha_j}  \left[ \nabla_a \omega^a + \nabla_\a \omega^\a\right]
 = 0 \ ,
\end{eqnarray}
having used the fact that $ \left( \prod_{j=1}^4 
\nabla_{\alpha_j} \right) \nabla_\a =0$ and $\left( \prod_{i=1}^4
\delta(i\nabla_{a_i}) \right) \nabla_a =0$ as a consequence of distributional properties discussed in \eqref{veB}. Notice that it is crucial 
that in the products we have all possible vector fields $\nabla_{\alpha_j}, \nabla_{a_i}$ 
to guarantee the invariance and Stokes' theorem. The extension of the present derivation 
to sub-supermanifolds is straightforward by using PCO's: suppose we have a $(n|m)$ supermanifold $\mathcal{SM}^{(n|m)}$ and a $(p|m), 0 \leq p \leq n$, sub-supermanifold $\mathcal{S}^{(p|m)}$. Given a $(p|m)$ form $\omega^{(p|m)}$ on the whole supermanifold $\mathcal{SM}^{(n|m)}$, we can integrate it on $\mathcal{S}^{(p|m)}$ as
\begin{equation}
	\int_{\mathcal{S}\hookrightarrow \mathcal{SM}} \omega^{(p|m)*} = \int_{\mathcal{SM}} \omega^{(p|m)} \wedge \mathbb{Y}^{(n-p|0)}_{\mathcal{S}} \equiv \mu \left[ \omega^{(p|m)} \wedge \mathbb{Y}^{(n-p|0)}_{\mathcal{S}} \right] \ ,
\end{equation}
where $\mathbb{Y}^{(n-p|0)}_{\mathcal{S}}$ is the Poincar\'e dual of the immersion $\mathcal{S}^{(p|m)}\hookrightarrow \mathcal{SM}^{(n|m)}$. If we now assume that $\omega^{(p|m)} = d \left( \eta^{(p-1|m)} \right)$ and that we can drop boundary terms, we have
\begin{equation}
	\mu \left[ d \left( \eta^{(p-1|m)} \right) \wedge \mathbb{Y}^{(n-p|0)}_{\mathcal{S}} \right] = \mu \left[ d \left( \eta^{(p-1|m)} \wedge \mathbb{Y}^{(n-p|0)}_{\mathcal{S}} \right) \right] = 0 \ ,
\end{equation}
as in \eqref{sto0A}, where we have used the closure of $\mathbb{Y}^{(n-p|0)}_{\mathcal{S}}$.

With these manipulations we want to emphasize that the algebraic description of integro-differential operators allows on the one hand to obtain quick proofs of known results and on the other hand to have a strong formalism to explore new directions. In particular, we have highlighted that the extension to supermanifolds is totally natural within this framework, since even and odd operators are treated on the same algebraic ground.

%
%
%
%
\section{What about boundaries?}

Let us start from (2.11) in the purely bosonic case:
\begin{eqnarray}
\label{veDboundary}
\Xi f = \epsilon^{a_1 \dots a_n} \delta(i\partial_{a_1}) \dots \delta(i\partial_{a_n})
f(x_1, \dots, x_n) = 
\int_{-\infty}^\infty dt_1 \dots dt_n  f(t_1, \dots, t_n) \ ,
\end{eqnarray}
and consider as well
\begin{eqnarray}
\label{veBboundary}
\delta(i \partial_a) \partial_a f(x_1, \dots, x_n) =0 \ .
\end{eqnarray}
The two previous results encode Stokes' theorem, if we are integrating on a manifold $\mathcal{M}$ with no boundary $\partial \mathcal{M} = \emptyset$, as discussed in the previous section. In particular, we have that the operator $\Xi$ defined in \eqref{veDboundary} defines an integration over $\mathcal{M} \cong \mathbb{R}^n$.

But what if $\partial \mathcal{M} \neq \emptyset$? We have to modify the operator $\Xi$ in \eqref{veDboundary} as follows:
\begin{equation}\label{BA}
	\tilde{\Xi} = \epsilon^{a_1 \dots a_n} \delta(i\partial_{a_1}) \dots \delta(i\partial_{a_n}) \left[ - \Theta \left( \partial \mathcal{M} \right) \right] \ ,
\end{equation}
where $\Theta$ is the Heaviside step function and the minus sign is for later convenience. If we include the form integration, we have
\begin{eqnarray}
\label{totAboundary}
\tilde{\Xi} \Xi^* &=& \Xi \, \epsilon_{m_1 \dots m_n} \delta(\iota_{\partial_{m_1}}) \dots  \delta(\iota_{\partial_{m_n}}) \left[ - \Theta \left( \partial \mathcal{M} \right) \right]  \nonumber \\
&=& 
 \Xi \, \epsilon_{m_1 \dots m_n} \iota_{\partial_{m_1}} \dots  \iota_{\partial_{m_n}} \left[ - \Theta \left( \partial \mathcal{M} \right) \right] \ .
\end{eqnarray}
Clearly, if $\partial \mathcal{M} = \emptyset$ we put $\Theta \left( \partial \mathcal{M} \right) =1$. This operator really encodes the Stokes' theorem for manifolds with non-trivial boundaries: given $\omega \in \Omega^{n-1} \left( \mathcal{M} \right)$, we have
\begin{eqnarray}
	\nonumber \tilde{\Xi} \Xi^* \left( d \omega \right) &=& \epsilon^{a_1 \dots a_n} \delta(i\partial_{a_1}) \dots \delta(i\partial_{a_n}) \epsilon_{m_1 \dots m_n} \delta(\iota_{\partial_{m_1}}) \dots  \delta(\iota_{\partial_{m_n}}) \left[ - \Theta \left( \partial \mathcal{M} \right) d \omega \right] = \\
	\nonumber &=& \epsilon^{a_1 \dots a_n} \delta(i\partial_{a_1}) \dots \delta(i\partial_{a_n}) \epsilon_{m_1 \dots m_n} \delta(\iota_{\partial_{m_1}}) \dots  \delta(\iota_{\partial_{m_n}}) \left[ - d \left( \Theta \left( \partial \mathcal{M} \right)  \omega \right) + \left( d \Theta \left( \partial \mathcal{M} \right) \right) \omega \right] = \\
	\nonumber &=& \epsilon^{a_1 \dots a_n} \delta(i\partial_{a_1}) \dots \delta(i\partial_{a_n}) \epsilon_{m_1 \dots m_n} \delta(\iota_{\partial_{m_1}}) \dots  \delta(\iota_{\partial_{m_n}}) \left[ \mathbb{Y}_{\partial \mathcal{M}} \omega \right] = \\
	&=& \label{Stokesboundary}  \int_{\partial \mathcal{M}} \omega \ ,
\end{eqnarray}
where we have used
\begin{equation}
	\epsilon^{a_1 \dots a_n} \delta(i\partial_{a_1}) \dots \delta(i\partial_{a_n}) \epsilon_{m_1 \dots m_n} \delta(\iota_{\partial_{m_1}}) \dots  \delta(\iota_{\partial_{m_n}}) \left[ d \left( \Theta \left( \partial \mathcal{M} \right)  \omega \right) \right] = 0 \ ,
\end{equation}
in analogy to \eqref{stoA}, since $\mathbb{R}^n$ has no boundary, and that $ d \Theta \left( \partial \mathcal{M} \right) = \mathbb{Y}_{\partial \mathcal{M}}$, i.e., the Poincar\'e dual that localizes on the boundary of $\mathcal{M}$.

Let us consider an easy example: we want to integrate $d f$, $f = f_x dx + f_y dy \in \Omega^1 \left( \mathbb{R}^2 \right)$ on the upper half plane $\mathcal{M} = \left\lbrace \left( x,y \right) | y \geq 0 \right\rbrace$. We have
\begin{equation}
	\partial \mathcal{M} = \left\lbrace \left( x,y \right) | y = 0 \right\rbrace \ \implies \ -\Theta \left( \partial \mathcal{M} \right) = \Theta \left( y \right) \ .
\end{equation}
By integrating $df$ we obtain
\begin{equation}
	\delta \left( i \partial_x \right) \delta \left( i \partial_y \right) \iota_x \iota_y \left[ \Theta \left( y \right) \left( - \partial_y f_x + \partial_x f_y \right) dx dy \right] = - \int_{- \infty}^\infty [dx] \int_{- \infty}^\infty [dy] \Theta (y) \partial_y f_x = $$ $$ = \int_{- \infty}^\infty [dx] \int_{- \infty}^\infty [dy] \delta (y) f_x = \int_{- \infty}^\infty [dx] \left. f_x \right|_{y = 0} = \int_{\partial \mathcal{M}} \left. f \right|_{\partial \mathcal{M}} \ .
\end{equation}
where we denoted by $[dx]$ and $[dy]$ the integration variables. 

Hence, as this example shows, the term $\Theta \left( \partial \mathcal{M} \right)$ in \eqref{BA} amounts to modify the integration from $\mathbb{R}^n$ to $\mathcal{M}$. 

The extension to supermanifolds does not require great modifications: the boundary of a supermanifold is the boundary of the reduced manifold (i.e., the embedded bosonic manifold defining the body) itself. In other words, consistent submanifolds are (locally) defined via bosonic constraints only. This means that the extension of \eqref{totAboundary} to supermanifolds lead to a modification of the operator $\mu$ defined in \eqref{totBA} as
\begin{equation}
	\mathcal{\mathlarger{\mathlarger \mu}} \ \longrightarrow \ \tilde{\mathcal{\mathlarger{\mathlarger \mu}}} = \tilde{\mathlarger{\mathlarger \Xi}} \mathlarger{\mathlarger \Xi^*}= \prod_{i=1}^n
\delta(i\nabla_{a_i}) \iota_{\nabla_{a_i}} \prod_{j=1}^m 
\nabla_{\alpha_j} \delta(i \iota_{\nabla_{\a_j}}) \left[ - \Theta \left( \partial \mathcal{M} \right) \right] \ .
\end{equation}
The extension of Stoke's theorem in the present formalism to supermanifolds follows directly from \eqref{Stokesboundary}, since the Berezin integration on $\theta$'s and the algebraic integration on $d \theta$'s are left untouched.

It would be interesting to investigate the case of submanifolds with non-trivial and non-maximal odd dimension: the task is far from trivial and is related to a class of forms which is not described in the present text, pseudoforms. We will not comment on this topic, the interested reader may refer to \cite{Witten:2012bg,Cremonini:2019aao} for an introduction and some applications of pseudoforms.

\section{The Integral Form of the Supergravity}

In this section, we apply the above formalism to $D=4$, $N=1$ supergravity. The result is well-known 
and it is used as a check on our formalism. Given the volume form ${\rm Vol}^{(4|4)}$ described in the 
previous sections and the measure $\mathlarger{\mathlarger \mu}$  on the 
tangent bundle $\Pi T\mathcal{SM}^{(4|4)}$, we can 
write the natural expression 
\begin{eqnarray}
\label{suagraA}
\mathlarger{\mathlarger \mu}\left[{\rm Vol}^{(4|4)} \right] &=& \prod_{i=1}^4 
\delta(\nabla_{a_i}) \nabla_{\alpha_i} \iota_{\nabla_{a_i}}  \delta(\iota_{\nabla_{\alpha_i}})
\Big[ \delta(E^{a_i}) \delta(E^{\alpha_i})  \Big] 
=
\int {\rm Ber}(\mathbb{E}) \ ,
\end{eqnarray}
which is the Berezin-Lebesgue integral over the coordinates $(x^m, \theta^\mu)$ of ${\rm Ber}(\mathbb{E})$. It is 
well-know how to compute explicitly this integral \cite{Wess:1992cp,GGRS,Buchbinder:1998qv}
and, after a long computation, one is able to see the full action of old minimal supergravity. 

A remark is in order: one can use the Hodge dual operator to build the action by integrating $\star 1$, as in general 
relativity for the cosmological constant term using the results of \cite{Catenacci:2016qzd,Castellani:2016ibp}. The star operation encodes the informations on the vielbeins, on the gravitinos and on the auxiliary fields.  

Is it possible to derive \eqref{suagraA} from the geometric approach to supergravity \cite{cube}? 
This approach encodes the equations of motion and the symmetries in the  
rheonomic Lagrangian $ {\cal L}^{(4|0)}(E,\varpi)$ which is a $(4|0)$ superform. As it stands, the rheonomic Lagrangian cannot be integrated on the entire supermanifold. 
However, we can to convert it to an integral form by multiplying it by a PCO 
${\mathbb Y}^{(0|4)}(E)$. The choice of the latter is not unique and, any other choice, differing by exact 
terms, does not change the action if $ {\cal L}^{(4|0)}(E,\varpi)$ is closed (which is the case for $N=1$, $D=4$ supergravity, see \cite{cube}). With these 
ingredients, following the prescriptions described above, we can set 
\begin{eqnarray}\label{APCOE}
	S_{sugra} =\mathlarger{\mathlarger \mu}\left[ { \mathcal L}^{(4|0)}( E, \varpi) \wedge  {\mathbb Y}^{(0|4)}(E) \right] \ .
	\end{eqnarray}
The choice of the representative of ${\mathbb Y}^{(0|4)}(E)$ determines which symmetries 
of the final actions are manifest, but, since $ { \mathcal L}^{(4|0)}( E,\varpi) \wedge  {\mathbb Y}^{(0|4)}(E)$ is a top integral form, 
it is automatically invariant under superdiffeomorphisms. In \cite{inprep-Leonardo-Antonio-Carlo}
a complete discussion for $D=4$ and $N=1$ will be given. On the other side, in \cite{Castellani:2016ibp} the $D=3$ and $N=1$ 
case is discussed in detail. 

Finally, one can build other observables and couple them the supergravity 
action to add other degrees of freedom. Thanks to the powerful formulation of integration on the tangent bundle, we 
have to build integral forms with the wanted dynamical variables by following the well-know strategy used in General Relativity. 
For example a possible candidate is $\tau^{(4|4)} = \Phi {\rm Vol}^{(4|4)}$, 
where $\Phi$ is an invariant scalar field (the construction of new actions based on these observations has been pursued in 
\cite{DelMonte:2016czb} and in \cite{Castellani:2016ibp}). The coupling to supergravity arises 
as 
\begin{eqnarray}
\label{paceA}
\mathlarger{\mathlarger \mu}\left[\Phi {\rm Vol}^{(4|4)} \right] = \int {\rm Ber}(\mathbb{E})\Phi(x,\theta) \ ,
\end{eqnarray}
which is a Berezin-Lebesgue integral. Of course, this is not the only possibility, any integral form $\omega^{(4|4)}$ is suitable for this purpose. 
Further analysis will be matter of investigations in \cite{inprep-Leonardo-Antonio-Carlo}. 

\section{The Integral Form of the Volkov-Akulov Theory}

In the present section, we consider another application of the formalism: Volkov-Akulov theory.\footnote{P.A.G. is grateful to D. Sorokin for suggesting this application.} It turns out that this application 
has two important features: on one side, it gives a geometrical interpretation of the Volkov-Akulov theory 
with non-linear symmetries; on the other side, it allows us to construct a new PCO representing 
a superembedding of a bosonic submanifold into a supermanifold. Let us illustrate this point first. 

As discussed in app. B, PCO's correspond to given embeddings of a submanifold into 
a supermanifold. This is locally described in terms of a set of constraints $\phi^\a(x,\theta) = 0$ which indicate 
how the fermionic coordinates are embedded. Generically, this is done by using a set of anticommuting superfields 
$\phi^\a(x,\theta)$ which can be expanded, without loss of generality, as 
\begin{eqnarray}
\label{VAA}
\phi^\a(x,\theta) =\theta^\a + \phi^\a_{[\b\g\delta]} (x)  \theta^\b  \theta^\g  \theta^\delta + \dots  =0 \ ,
\end{eqnarray}
where $\phi^\a_{[\b\g\delta]} (x)$ are conventional commuting functions of the spacetime coordinates $x^m$. 
The ellipsis stand for additional pieces, when there are more than four $\theta$'s (see also eqs. 
\eqref{APPC}-\eqref{APPF}). These equations are constraints on $\theta^\a$ and the corresponding PCO 
can be easily written as $\displaystyle \prod \phi^\alpha \delta \left( d \phi^\alpha \right) $. However, in quantum field theory there exist also anticommuting functions 
such as for example the Dirac field, the gluinos and the Goldstino. Therefore, we can set a more 
generic expression for \eqref{VAA} as 
\begin{eqnarray}
\label{VAB}
\phi^\a(x,\theta) =\Lambda^\a(x) - \theta^\a + \phi^\a_{\b\g}(x)\theta^\b \theta^\gamma+ \phi^\a_{[\b\g\delta]} (x)  \theta^\b  \theta^\g  \theta^\delta   =0 \ ,
\end{eqnarray}
where $\Lambda^\a(x), \phi^\a_{\b\g}(x)$ are anticommuting functions. The normalization of the $\theta^\a$ term is 
chosen to simplify the expression. In particular, if we set $ \phi^\a_{\b\g}(x)=\phi^\a_{[\b\g\delta]} (x)=0$, we can identify 
$\Lambda^\a(x)$ with the Goldstino field. This means that the embedding of the bosonic submanifold is chosen in terms 
of the Goldstino field, (see \cite{Sorokin:1999jx} and references therein for an exhaustive discussion) which is then easily implemented in the present framework. 
We can define a new PCO
\begin{eqnarray}
\label{AKA}
 {\mathbb Y}^{(0|4)}(\theta, d\theta, \Lambda, d\Lambda) &=& (\theta^1 - \Lambda^1) \dots (\theta^4 - \Lambda^4) 
 \delta(d\theta^1 - d\Lambda^1) \dots 
 \delta(d\theta^4 - d\Lambda^4)  \nonumber \\
&=& \prod_{\a=1}^4  (\theta^\a - \Lambda^\a) \delta(d\theta^\a - d\Lambda^\a) \nonumber \\
&=& e^{- {\cal L}_\Lambda} \prod_{\a=1}^4  \theta^\a \delta(d\theta^\a) \ ,
\end{eqnarray}
which is trivially closed because of the Dirac delta properties. In the last line, we used the Lie derivative ${\cal L}_\Lambda$ 
along the commuting vector field $\Lambda = \Lambda^\a D_\a$ (with $D_\a$ the superderivative), to generate the shifted 
PCO w.r.t. to the simple PCO (see App. B). The same strategy has been adopted also in \cite{Cremonini:2020mrk} in the context of supersymmetric Wilson Operators. 

The exterior derivative of the Goldstino  $d\Lambda^\a = dx^a \partial_a \Lambda^\a$ is written in terms of
an anticommuting rectangular matrix $\partial_a \Lambda^\a$, which is the usual ingredient in the Volkov-Akulov action. 
The arguments of $ {\mathbb Y}^{(0|4)}(\theta,d\theta, \Lambda, d\Lambda)$ remind us the dependence on the flat gravitinos 
$\psi^\a=d\theta^\a$ and the Goldstino field $\Lambda^\a$. The presence of naked $\theta$'s implies that this is not 
invariant under supersymmetry variations.
%
%
The variation of the Goldstino 
field $\Lambda$ inside the PCO leads to (recall that any variation of a PCO is $d$-exact) 
\begin{eqnarray}
\label{AKAB}
\delta_{\Lambda}  {\mathbb Y}^{(0|m)}(\theta, \Lambda)  = d \Big[ 
 \prod_{\a=1}^m  (\theta^\a - \Lambda^\a) 
\delta \Lambda^\beta \iota_\beta  \prod_{\a=1}^m  \delta(d\theta^\a - d\Lambda^\a) 
\Big] \ .
\end{eqnarray}

The action of a suitable integral form for flat Volkov-Akulov action is now written as
\begin{eqnarray}
\label{AKB}
S_{V-A} = \mathlarger{\mathlarger \mu}\left[   \epsilon_{a_1 \dots a_4}  V^{a_1}\wedge \dots \wedge V^{a_4}\wedge 
 {\mathbb Y}^{(0|4)}(\theta,d\theta, \Lambda, d\Lambda) \right] \ ,
\end{eqnarray}
where we introduced the flat supervielbeins $V$ as in \eqref{exDA}. To check that \eqref{AKB} produces the correct action, we observe from \eqref{AKA} that $\theta$ is 
substituted with the Goldstino $\Lambda$ and the gravitino $d\theta$ is substituted with the differential of the Goldstino 
$d\Lambda$. By substituting in the expression for the flat supervielbeins one gets 
\begin{eqnarray}
\label{AKC}
S_{V-A} &=& \mu \left[  \epsilon_{a_1 \dots a_4}  (dx^{a_1} + \Lambda \gamma^{a_1} dx^b \partial_b \Lambda) 
\wedge \dots \wedge  (dx^{a_4} + \Lambda \gamma^{a_4} dx^c \partial_c \Lambda)  \wedge 
{\mathbb Y}^{(0|4)} \right] \nonumber \\
&=& \mu \left[ {\rm det}( \delta_a^b + \Lambda \gamma^b \partial_a \Lambda) \epsilon_{a_1 \dots a_4} dx^{a_1} \wedge 
dx^{a_4} \prod_{\a=1}^4 \widetilde{\theta}^\a
\delta(d\widetilde{\theta}^\a) \right] \nonumber \\
&=& \int  {\rm det}( \delta_a^b + \Lambda \gamma^b \partial_a \Lambda) \ ,
\end{eqnarray}
where we have renamed $ \widetilde{\theta}^\a = \theta^\a - \Lambda^\a$ and $\widetilde{\Lambda}^\a = 
d\theta^\a - d\Lambda^\a$. 
The integration over $\widetilde{d\theta}$ is done by using the Dirac delta's, while the integration 
over $ \widetilde{\theta}$ is the usual Berezin integral, as largely discussed n the previous sections. Finally, the remaining integration is the usual 
integration on the submanifold ${\cal M}^4$ of the supermanifold ${\cal SM}^{(4|4)}$ and 
the last expression is the usual Volkov-Akulov action \cite{VA}.  

In the present case the superform Lagrangian ${\cal L}^{(4|0)}(V) =  \epsilon_{a_1 \dots a_4}
V^{a_1}\wedge \dots \wedge V^{a_4}$ contains only the geometrical data of the supermanifold 
and it does not contain the dynamical fields. In addition, ${\cal L}^{(4|0)}(V)$ is not closed:
\begin{eqnarray}
\label{AKD}
d{\cal L}^{(4|0)}(V) = 4 \epsilon_{a_1 \dots a_4} d\theta \gamma^{a_1} d\theta \wedge V^{a_2}\wedge  \dots \wedge V^{a_4}  \neq 0 \ .
\end{eqnarray}

However, the non-closure of ${\cal L}^{(4|0)}$ is crucial in this case 
otherwise any change of the dynamical field $\delta \Lambda$ would vanish since it would correspond to an exact variation of the PCO: 
 \begin{eqnarray}
\label{AKF}
\delta_{\Lambda}  {\mathbb Y}^{(0|4)} = d \eta^{(-1|4)} \ ,
\end{eqnarray}
under a generic variation of $\Lambda$. 

Within this framework, we can now build the coupling of the Volkov-Akulov 
action to dynamical supergravity fields by promoting the flat construction of the above paragraphs. 
First of all, we need to construct a corresponding PCO in terms of the dynamical fields $(E^a, E^\a)$. 
For that, we use the Euler vector $\mathcal{E} = \theta^\mu \partial_\mu + f^m(x, \theta) \partial_m = \Theta^\alpha \nabla_\alpha$, defined in 
app. B, satisfying 
\begin{eqnarray}
\label{SUEA}
\iota_\mathcal{E} E^\a = \Theta^\a\,, ~~~~
\iota_\mathcal{E} E^a = 0\,. ~~~~
\end{eqnarray}
The covariant derivative of $\Theta^\a$ reads
\begin{eqnarray}
\label{SUEB}
 \nabla \Theta^\a = \nabla ( \iota_\mathcal{E} E^\a)   = E^\a - \iota_\mathcal{E} T^\a  + \Omega^\a_{~\b} E^\b\, ,
\end{eqnarray}
so that we can write the curved PCO in \eqref{totCAA} as
\begin{eqnarray}
\label{SUEC}
{\mathbb Y}^{(0|4)} = \prod_{\a =1}^{4} \Theta^\alpha \delta \left( \nabla \Theta^\alpha \right) = \prod_{\a =1}^{4}\Big(\iota_\mathcal{E} E^\a \Big) 
\delta\Big((1+ \Omega)^\a_{~\b} E^\b- \iota_\mathcal{E} T^\a\Big)\, ,
\end{eqnarray}
which is closed and not exact. This expression can be simplified so that the argument of the Dirac delta reduces to $E^\a$, since the other terms are all proportional to $\Theta$ and are then annihilated by the four $\Theta$'s in front of the delta's, as commented in App. B. 
We can finally rewrite the PCO by shifting it with the Goldstino 
field $\Lambda = \Lambda^\a(x)\nabla_\a$, in analogy to \eqref{AKA}, so that we can construct the Volkov-Akulov action 
coupled to dynamical fields $(E^\a, E^a)$ as
\begin{eqnarray}
\label{SUED}
S_{V-A} =\mathlarger{\mathlarger \mu}
\left[   
\epsilon_{a_1 \dots a_4} E^{a_1} \wedge \dots \wedge E^{a_4}  e^{- {\cal L}_\Lambda}
\Big( \prod_{\a =1}^{4} \iota_\mathcal{E} E^\a \delta(E^\a) \Big) 
\right]\,. 
\end{eqnarray}
This action is manifestly invariant under superdiffeomorphisms on the entire 
supermanifold ${\cal SM}^{(4|4)}$ since it is a top-form. By freezing the dynamical fields 
$(E^a, E^\a)$ to the rigid case, the action reduces to Volkov-Akulov action in $D=4$ for the Goldstino 
field $\Lambda^\a(x)$. By setting the Goldstino to zero, we get the action 
 \begin{eqnarray}
\label{SUD}
S_{V-A}(\Lambda =0) = \int_{{\cal SM}^{(4|4)}} 
\Big( \epsilon_{a_1 \dots a_4} E^{a_1} \wedge \dots \wedge E^{a_4} \Big) 
\Big( \prod_{\a =1}^{4} \iota_\mathcal{E} E^\a \delta(E^\a) \Big)\,. 
\end{eqnarray}
\eqref{SUD} corresponds to a cosmological constant term which is always present in the 
Volkov-Akulov theory, when the Goldstino field is set to zero. 
In the D=3 case (see \cite{Castellani:2016ibp}) this indeed leads to the correct cosmological 
term and the gravitino mass, however in D=4 it happens that a suitable choice of 
$\mathcal{E}$ reduces the expression to usual superspace chiral densities (see \cite{Wess:1992cp,Buchbinder:1998qv}). 
That will be matter of a forthcoming publication \cite{inprep-Leonardo-Antonio-Carlo}. 

\section{Perspectives}

We hope we have convinced the reader of the powerfulness of integral forms to formulate 
supergravity theory in a more geometrical way. We show two examples, namely a compact 
formula for supergravity $D=4, N=1$ and a new derivation of the Volkov-Alkulov theory with 
its couplings to supergravity background. Future directions are the extension of the 
present technique to higher dimensional cases and extended supersymmetry and the final goal is 
the 
application to path integrals and quantum field theory.

\section*{Acknowledgements}
\noindent This work has been partially supported by Universit\`a del Piemonte Orientale research funds. We thank L. Castellani, R. Catenacci, S. Noja, S.Penati and D. Sorokin for many useful discussions. 

\appendix

\section{A Brief Review on Integral Forms}

In this appendix we want to recall the main definitions and computation techniques for integral forms. For a more exhaustive review or for a more rigorous approach to integral forms we suggest \cite{Belo,Witten:2012bg,Catenacci:2010cs,Cremonini:2019aao}.

We consider a supermanifold ${\cal SM}^{(n|m)}$ with $n$ bosonic and $m$ fermionic dimensions. We denote the local coordinates in an open set as $(x^a, \theta^\alpha), a=1,\ldots,n , \alpha=1,\ldots,m$. A generic $(p|0)$-form, i.e., a \emph{superform}, has the following local expression
\begin{equation}\label{ABRIFA}
	\omega^{(p|0)} = \omega_{[i_1 \ldots i_r](\alpha_1 \ldots \alpha_s)} \left( x , \theta \right) dx^{i_1} \wedge \ldots \wedge dx^{i_r} \wedge d \theta^{\alpha_1} \wedge \ldots \wedge d \theta^{\alpha_s} \ , \ p=r+s \ .
\end{equation}
The coefficients $\omega_{[i_1 \ldots i_r](\alpha_1 \ldots \alpha_s)}(x,\theta)$ are a set of superfields and the indices $a_1 \dots a_r$, $\alpha_1 \dots \alpha_s$ are anti-symmetrized and symmetrised, respectively, because of the rules (we omit the \virgolette $\wedge$" symbol)
\begin{equation}\label{ABRIFB}
	dx^i dx^j = - dx^j dx^i \ , \ d \theta^\alpha d \theta^\beta = d \theta^\beta d \theta^\alpha \ , \ dx^i d \theta^\alpha = d \theta^\alpha d x^i \ .
\end{equation}
Namely, we assign \emph{parity} $1$ to odd forms and $0$ to even forms:
\begin{equation}\label{ABRIFC}
	\left| dx \right| = 1 \ , \ \left| d \theta \right| = 0  \ .
\end{equation}
Since superforms are generated both by commuting and anti-commuting forms, we immediately see that there is no top form. 
In other words, if one looks for the analogous of the determinant bundle on a supermanifold, one has to consider a different space of forms, namely the \emph{integral forms}. A generic integral form locally reads 
\begin{equation}\label{ABRIFD}
	\omega^{(p|m)} = \omega_{[i_1 \ldots i_r]}^{(\alpha_1 \ldots \alpha_s)} \left( x , \theta \right) dx^{i_1} \wedge \ldots \wedge dx^{i_r} \wedge \iota_{\alpha_1} \ldots \iota_{\alpha_s} \delta \left( d \theta^1 \right) \wedge \ldots \wedge \delta \left( d \theta^m \right) \ ,
\end{equation}
where $\delta \left( d \theta \right)$ is a (formal) Dirac delta function and $\iota_\alpha$ denotes the interior product. The integration on $d\theta$'s is defined 
algebraically by setting 
\begin{equation}\label{ABRIFE}
	\int_{d \theta} \delta \left( d \theta \right) = 1 \,, ~~~ \int_{d \theta} f(d\theta) \delta \left( d \theta \right) = f(0)  \ ,
\end{equation}	
for a generic test function $f(d\theta)$. 	
The symbol  $\delta \left( d \theta \right)$ satisfies the usual distributional equations 
\begin{equation}\label{ABRIFEA}
	 \ \ d \theta \delta \left( d \theta \right) = 0 \,,  
	  \delta \left( \lambda d \theta \right) = \frac{1}{\lambda} \delta \left( d \theta \right) \,, 
	  d \theta  \delta^{(1)} \left( d \theta \right) = - \delta \left( d \theta \right) \,, 
	  d \theta  \delta^{(p)} \left( d \theta \right) = - p \delta^{(p-1)} \left( d \theta \right) \,, 
\end{equation}	  
We sometimes denote by $\iota_\alpha \delta(d\theta^\a) \equiv \delta^{(1)}(d\theta^\a)$. Additional properties are 
 \begin{equation}\label{ABRIFEAA}
	 \delta \Big( d \theta^\alpha \Big) \wedge \delta \left( d \theta^\beta \right) = - \delta \left( d \theta^\beta \right) \wedge \delta \Big( d \theta^\alpha \Big) \ \ , \ \ dx \wedge \delta \left( d \theta \right) = - \delta \left( d \theta \right) \wedge d x \ ,
	 \end{equation}
indicating that actually these are not conventional distributions, but rather \emph{de Rham currents}. 

Given these properties, we retrieve  a  top form among integral forms as
\begin{equation}\label{ABRIFF}
	\omega_{top}^{(n|m)} = \omega \left( x , \theta \right) dx^1 \wedge \ldots \wedge dx^n \wedge \delta \left( d \theta^1 \right) \wedge \ldots \wedge \delta \left( d \theta^m \right) \ ,
\end{equation}
where $\omega \left( x , \theta \right)$ is a superfield. The space of $(n|m)$ forms corresponds to 
the \emph{Berezinian bundle} since  the generator $dx^1 \wedge \ldots \wedge dx^n \wedge \delta \left( d \theta^1 \right) \wedge \ldots \wedge \delta \left( d \theta^m \right)$ transforms as the superdeterminant of the Jacobian.

One can also consider a third class of forms, with non-maximal and non-zero number of delta's: the \emph{pseudoforms}. A general pseudoform with $q$ Dirac delta's is locally given by
\begin{equation}\label{ABRIFG}
	\omega^{(p|q)} = \omega_{[a_1 \ldots a_r](\alpha_1 \ldots \alpha_s)[\beta_1 \ldots \beta_q]} \left( x , \theta \right) dx^{a_1} \wedge \ldots \wedge dx^{a_r} \wedge d \theta^{\alpha_1} \wedge \ldots \wedge d \theta^{\alpha_s} \wedge \delta^{(t_1)} \left( d \theta^{\beta_1} \right) \wedge \ldots \wedge \delta^{(t_q)} \left( d \theta^{\beta_q} \right) \ ,
\end{equation}
where $\delta^{(i)} \left( d \theta \right) \equiv \left( \iota \right)^i \delta \left( d \theta \right)$. The form number is obtained as 
\begin{equation}\label{ABRIFH}
	p = r + s - \sum_{i=1}^q t_i \ ,
\end{equation}
since the contractions carry negative form number. The two quantum numbers $p$ and $q$ in eq. \eqref{ABRIFH} correspond to the {\it form} number and the {\it picture} number, respectively, and they range as $-\infty < p < +\infty$ and $0 \leq q \leq m$, so the picture number counts the number of delta's. If $q=0$ we have superforms, if $q=m$ we have integral forms, if $0<q<m$ we have pseudoforms.

As in conventional geometry, we can define the integral of a top form on a supermanifold (more rigorously, the integration 
is on  the parity-shifted tangent space $\Pi T\mathcal{SM}$) as
 \begin{eqnarray}\label{ABRIFI}
I[\omega] = \int_{{\cal SM}} \omega_{top}^{(n|m)} =  \int_{x, \theta} \omega(x, \theta)\ ,
\end{eqnarray}
where we integrated over the odd variables $dx$ and over the even variables $d \theta$ 
to obtain an ordinary superspace integral over the variables $(x, 
\theta)$. Stokes' theorem for integral forms applies, as we largely discuss in the text. 

\section{Picture Changing Operators}

The strategy that we use for constructing integral forms and the corresponding supermanifold integrals is the following. Given a $(p|0)$-superform $\omega^{(p|0)}$ on a supermanifold ${\cal SM}$ of dimension $(n|m)$ ($n \geq p$), its integration over a $p$-dimensional bosonic submanifold ${\cal N} \subset  {\cal SM}$ can be defined  as the integration on the entire supermanifold of the integral form $\omega^{(p|0)} \wedge \mathbb{Y}^{(n-p|m)}_{\cal N}$, where $\mathbb{Y}^{(n-p|m)}_{\cal N}$ is the \emph{Poincar\'e dual} to the immersion of ${\cal N}$ into ${\cal SM}$ \cite{grassi-castellani}. Precisely, if we denote $\omega^{(p|0)*} \equiv i^* \omega^{(p|0)}$, where $\displaystyle i : \mathcal{N} \hookrightarrow \mathcal{SM} $ is the  immersion of ${\cal N}$ into ${\cal SM}$ \footnote{Precisely, we consider ${\cal N} \subset {\cal M}$ where ${\cal M}$ is the bosonic component of  $ {\cal SM}$ known in the literature as the body.}, we define
\begin{eqnarray}\label{APCOA}
	\int_{\cal N} \omega^{(p|0)*} = \int_{\cal SM} \omega^{(p|0)} \wedge \mathbb{Y}^{(n-p|m)}_{\cal N} \ .
\end{eqnarray}
The second expression is the integral over the whole supermanifold of a $(n|m)$-dimensional top form to which we can then apply the usual Cartan calculus rules.
Operator $\mathbb{Y}^{(n-p|m)}_{\cal N}$ is also known as {\it Picture Changing Operator} (PCO), being related to a similar concept in string theory. 

The PCO in \eqref{APCOA} is independent of the coordinates, it only depends on the immersion through its homology class. Its main properties are:
\begin{equation}\label{APCOB}
	d{\mathbb Y}^{(n-p|m)}_{\cal N} = 0 \ \ , \ \ \mathbb{Y}^{(n-p|m)}_{\cal N} \neq d \Sigma^{(n-p-1|m)} \ ,
\end{equation}
and by changing the immersion $i$ to an homologically equivalent surface ${\cal N}'$, the new Poincar\'e dual $\mathbb{Y}^{(n-p|m)}_{\cal N'}$  differs from the original one by $d$-exact terms:
\begin{equation}\label{APCOC}
	\delta {\mathbb Y}^{(n-p|m)}_{\cal N} = {\mathbb Y}^{(n-p|m)}_{\cal N'} - {\mathbb Y}^{(n-p|m)}_{\cal N} = d \Lambda^{(n-p-1|m)} \ .
\end{equation} 
As a consequence of this, if $\omega^{(p|0)}$ is a closed form, then \eqref{APCOA} is automatically invariant under any change of the embedding (modulo boundary contributions). 

It is worth discussing some general properties of ${\mathbb Y}^{(4|4)}$ used in the text (see \eqref{totC} and \eqref{totCAA}). 
We recall that, since $ {\mathbb Y}^{(4|4)}$ is closed, we have
\begin{eqnarray}
\label{APPA}
{\mathcal L}_{X} {\mathbb Y}^{(4|4)} = d \left( \iota_X {\mathbb Y}^{(4|4)} \right) \ ,
\end{eqnarray}
for a generic vector field 
\begin{eqnarray}
\label{APPB}
X &=& X^m \partial_m + X^\mu \partial_\mu = X^m (E^a_m \nabla_a + E^\a_m \nabla_\a) + 
X^\mu (E^a_\mu \nabla_a + E^\a_\mu \nabla_\a) \nonumber \\&=& X^a \nabla_a + X^\alpha \nabla_\a \ .
\end{eqnarray}
In the case of \eqref{totC}, it is easy to show that it is invariant under the reparametrization 
\begin{eqnarray}
\label{APPC}
\theta^\mu \rightarrow f^\mu(x, \theta) \ .
\end{eqnarray}
In particular, since the functions $f^\mu(x, \theta)$ have to be fermionic, we can expand them into powers of $\theta$'s and 
therefore $f^\mu(x, \theta) = \partial_\nu f^\mu(x) \theta^\nu +  
\partial_\nu \partial_\sigma \partial_\rho f^\mu(x) \theta^\nu \theta^\sigma \theta^\rho + \dots$ 
then we have 
\begin{eqnarray}
\label{APPD}
\prod_{\mu=1}^4 \theta^\mu \rightarrow \det(  \partial_\nu f^\mu(x) ) \prod_{\mu=1}^4 \theta^\mu\,.
\end{eqnarray}
On the other side we have 
\begin{eqnarray}
\label{APPE}
\delta(d\theta^\mu) \rightarrow &&\delta(\partial_\nu f^\mu d\theta^\nu + dx^m \theta^\nu \partial_\sigma \partial_m f^\mu + \dots)  =
\delta(\partial_\nu f^\mu d\theta^\nu) + (dx^m \theta^\nu \partial_\sigma \partial_m f^\mu) \iota_\mu \delta(d\theta^\mu) + \dots \nonumber
\end{eqnarray}
where we Taylor-expanded the expression. By putting together all the $\delta$'s, we get 
\begin{eqnarray}
\label{APPF}
\prod_{\mu=1}^4 \delta \left( d\theta^\mu \right) \rightarrow \frac{1}{\det(  \partial_\nu f^\mu(x) )} \prod_{\mu=1}^4 \delta \left( d\theta^\mu \right) + \ldots \,,
\end{eqnarray}
where the ellipses stand for other terms which are at least proportional to $\theta$. By combining \eqref{APPE} and \eqref{APPF}, we immediately see that all such terms vanish, and then ${\mathbb Y}^{(4|4)}$ is invariant under transformations \eqref{APPC}. This is not true for conventional supersymmetric theories, since in that case the functions $f^\mu$ is fermionic without being $\theta$-dependent. 

The second expression in \eqref{totCAA} is defined in terms of the special Euler vector 
\begin{eqnarray}
\label{APPG}
&&{\mathcal E} = \theta^\mu \partial_\mu + f^m(x,\theta) \partial_m =  X^a \nabla_a + \Theta^\alpha \nabla_\a \ , \nonumber \\ 
 &&\Theta^\a = \theta^\mu E^\a_\mu + f^m(x,\theta) E^\a_m \ , ~~~~~ 
  X^a = \theta^\mu E^a_\mu + f^m(x,\theta) E^a_m \ ,
\end{eqnarray}
where the combinations $(X^a, \Theta^\a)$ are the new coordinates with flat indices. We can set $X^a=0$  by 
choosing the function $f^m(x,\theta) = E^m_\alpha \Theta^\a$, as can be verified by using \eqref{exN}. Then, we have 
\begin{eqnarray}
\label{APPH}
\iota_{\mathcal E} E^\a = \Theta^\a\,, ~~~
\iota_{\mathcal E} E^a = 0\,. ~~~
\end{eqnarray}
Applying the covariant differential $\nabla$ on $\Theta^\a$, we get
\begin{eqnarray}
\label{APPK}
\nabla \Theta^\a = \nabla \iota_{\mathcal E} E^\a = E^\a - \iota_{\mathcal E} T^\a + \Omega^\a_{~\b} E^\b
~~~~\Longrightarrow ~~~~
E^\a = [(1 + \Omega)^{-1}]^\a_{~\beta}(\nabla \Theta^\b  +  \iota_{\mathcal E} T^\b) \ ,
\end{eqnarray}
where $T^\a$ is the spinorial component of the supertorsion and $\Omega^\a_{~\b}= \iota_{\mathcal{E}} \varpi_{ab} 
(\gamma^{ab})^\a_{~\b}$ is a gauge parameter built in terms of the spin connection $\varpi$.

 In general, $\iota_{\mathcal E} T^\a$ does not 
vanish for dynamical supergravity fields and $\Omega^\a_{~\b} = 
 \varpi_{ab, \rho} \iota_{\mathcal E} E^\rho (\gamma^{ab})^\a_{~\b}$ .  Notice that $\Omega^\a_{~\b}$ is proportional to $\Theta$, hence it could be dropped in the expression of the curved PCO because of the product of the four $\Theta$'s in front of the delta's.
By using the expression in \eqref{APPK}, the curved PCO reads
\begin{eqnarray}
\label{APPKA}
{\mathbb Y}^{(0|4)} = \prod_{\a=1}^4 \Theta^\a \delta(\nabla \Theta^\a)  =\prod_{\a=1}^4 \iota_{\mathcal E} E^\a 
 \delta\Big( (1 + \Omega)^\a_{~\beta} E^\a - \iota_{\mathcal E} T^\a\Big) \ ;
\end{eqnarray}
it is closed because $\nabla^2 \Theta^\a = R_{ab} (\gamma^{ab})^\a_{~\beta} \Theta^\b$, 
since the indices on the $\Theta$'s are the flat Lorentz indices 
and because of the product of all $\Theta^\a$ in front of the delta's. Moreover, since $\iota_{\mathcal E} T^\a = \Theta^\b E^b T_{\b b}^{~~\a}$, i.e., it is proportional to $\Theta$,
${\mathbb Y}^{(0|4)}$ 
can be reduced to 
\begin{eqnarray}
\label{APPKAB}
{\mathbb Y}^{(0|4)} = \prod_{\a=1}^4 \Theta^\a \delta(\nabla \Theta^\a)  =\prod_{\a=1}^4 \Theta^\a 
 \delta( E^\a) \ .
\end{eqnarray}
As a last remark, the PCO's \eqref{totC} and \eqref{totCAA} and ${\mathbb Y}^{(0|4)}$ are not 
manifestly supersymmetric.

\end{document}